\documentclass[prl,aps,twocolumn]{revtex4}
\usepackage{amsmath}
\usepackage[english]{babel}
\usepackage{graphicx}

\newcommand{\bmx}{\begin{pmatrix}}
\newcommand{\emx}{\end{pmatrix}}
\newcommand{\bcs}{\begin{cases}}
\newcommand{\ecs}{\end{cases}}
\newcommand{\bqs}{\begin{eqnarray}}
\newcommand{\eqs}{\end{eqnarray}}
\newcommand{\s}{\sigma}
\newcommand{\pf}{\mathrm{Pf\phantom{,}}}

\begin{document}

\title{Anomalous Josephson current via
  Majorana bound states in topological insulators}
\author{P. A. Ioselevich$^{1,2}$ and M. V. Feigel'man$^{1,2}$} 
\affiliation{ 
$^{1}$L. D. Landau Institute for Theoretical Physics, Kosygin str.2, Moscow 119334, Russia and\\ 
$^{2}$Moscow Institute of Physics and Technology, Moscow 141700, Russia}

\begin{abstract}
We propose a setup involving Majorana bound states (MBS) hosted by a
vortex on a superconducting surface of a 3D Topological Insulator
(TI). We consider a narrow channel drilled across a TI slab with both sides covered by s-wave
superconductor. In the presence of a vortex pinned to such a channel, it
acts as a ballistic nanowire connecting the superconducting
surfaces, with a pair of MBS localized in it. The energies of the MBS possess
a $4\pi$-periodic dependence on the superconductive phase 
difference $\varphi$ between the surfaces. It results in the appearence
of an anomalous term in the current-phase relation, $I_a(\varphi)$ 
 for the supercurrent flowing along the channel between the
 superconductive surfaces. We have calculated the shape of the
 $4\pi$-periodic function $I_a(\varphi)$, as well as the  dependence  of its 
amplitude  on temperature and system parameters.
\end{abstract}

\maketitle

Since Majorana bound states (MBS) were predicted
to exist in solid state systems, a number of different suggestions on how
to detect them has been made. MBS  are predicted to exist in  systems
 characterized by both strong spin-orbit coupling and
 superconductivity. Examples include the surface of a topological
 insulator (TI)  covered by s-wave superconductor with vortices
 \cite{FuKane2008} or electrostatic defects \cite{MajoInDefects}, as
 well as ordinary semiconductor
 nanowires with spin-orbit coupling and proximity-induced
 superconductivity in the  presence of a sufficiently strong Zeeman
 field \cite{SemiNanowire}.
 While there are numerous suggestions on detection of
 MBS \cite{FuKane2008,Lee2009,BeenakkerAkhmerovNilsson2009,Sarma1,Sarma2},
 no experimental success has been reported yet. 
 One more way to track down MBS in a superconducting proximity system is
 to observe an anomalous $4\pi$-periodic phase
 dependence of a supercurrent. It was  shown by A.Kitaev~\cite{Kitaev2000},
 that a specific 1-D fermionic chain hosting a pair of MBS switches its ground state fermionic
 parity under an adiabatic change of the superconducting
 phase difference $\varphi$  by $2\pi$. 
 Thus, if the chain   conserves fermionic parity, its behavior has
 to be $4\pi$-periodic.

In the present letter we propose and study a system with rather
simple geometry, based on a TI sample covered by a
superconducting film, and calculate the anomalous $4\pi$-periodic
supercurrent it carries.
 More specifically, we consider a flat thin slab  of strong TI 
 with both its surfaces  covered by an s-wave
 superconducting film.  The slab, together with the
 superconductive films,  is pierced by a cylindric hole of radius $R$,
see Fig.1.  A single superconductive vortex with a flux quantum 
$\Phi_0 = hc/2e$ is pinned to that hole.
 The two superconducting surfaces  are connected far away from the hole,
 forming an SNS circuit with the hole acting as a normal
 region.  Below we demonstrate that MBS should be present  in this
 setup, and  calculate the anomalous $4\pi$-periodic  component
 $I_a(\varphi)$ of  the current flowing along the cylindric hole between the
 superconductive surfaces. 
 Related issues were considered in a recent preprint
~\cite{Hopf} in terms of the Hopf invariant and its relation to fermionic parity.
Note that in the SNS-like setup we consider, a Hopf invariant cannot be defined since a part of the
TI surface is not gapped. 
 
\begin{figure}
	\centering
		\includegraphics[width=0.40\textwidth]{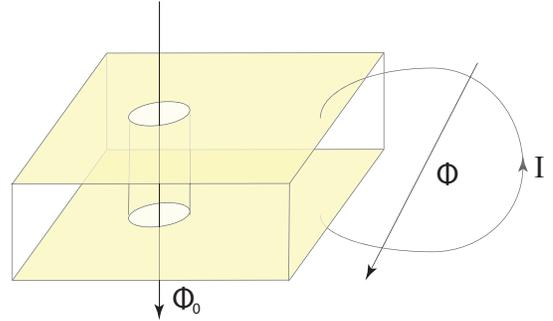}
	\caption{The system. A layer of TI has both surfaces covered
          by superconductor. 
A hole in the layer hosting a vortex forms an SNS-junction between the
surfaces. The superconducting
 surfaces close away from the hole, completing an SNS-circuit with
 supercurrent flowing through the hole.
	}\label{fig:model}
\end{figure}

In the simplest case of strong TI,  realized in  $Bi_2Se_3$ and
$Bi_2Te_3$, surface electrons are described by a single Dirac cone with
the Hamiltonian  $H=v_f\sigma\cdot\mathbf{p}-E_f$, where  $\sigma$ is
the spin operator and $E_f$ is the Fermi energy of the surface states of the TI.
 Below we consider the semiclassical limit $\Delta \ll E_f$ and put $\hbar=1$.
For the upper and lower surfaces  covered by s-wave superconductor, a pairing term 
$\Delta\psi^\dagger\psi^\dagger+h.c.$ arises due to proximity effect
~\cite{FuKane2008}:
\bqs
\hat{H}=(v_f\sigma\cdot \mathbf{p} - E_f)\tau_z + 
\Delta(\mathbf{r})(\tau_x\cos\varphi(\mathbf{r})+
\tau_y\sin\varphi(\mathbf{r}))
\label{h1}
\eqs 
The Pauli matrices $\mathbf{\tau}$ act in the Nambu-Gor'kov space,
$\hat{K}$ denotes complex conjugation. For the Hamiltonians $H_{u,l}$ 
acting on upper and lower surfaces, one should replace
$\Delta (\mathbf{r})$ and $\varphi(\mathbf{r})$ in Eq.(\ref{h1})  with $\Delta_{u,l}(\mathbf{r})$ and $\varphi_{u,l}(\mathbf{r})$ 
correspondingly.
The operator $\hat{H}$ anticommutes with the electron-hole conjugation
operator $\Xi=\sigma_y\tau_y\hat{K}$; we emphasize that this property
holds for any Bogolyubov-De Gennes Hamiltonian with a most general form of
single-electron spectrum, including terms that break time-inversion symmetry.
The symmetry $\{\hat{H},\hat{\Xi}\}=0$ divides the eigenstates of the
Hamiltonian \eqref{h1} into a set of conjugate pairs
$\psi_E,\psi_{-E}=\Xi\psi_E$ with opposite energies $(E,-E)$ and,
possibly, a number of self-conjugate states with zero energy. 
In the  basis of self-conjugated  states all matrix elements of
\eqref{h1} are purely  imaginary. The Hermitian matrix $H$ is
antisymmetric in this  basis, therefore the corresponding Pfaffian $\pf H$ can be defined. 
It is easy to see that \textit{generically}  the existence of $\pf H$ protects twofold
 degenerate  zero-energy levels of $H$  against  splitting under adiabatic variations  of
$H$. Indeed, consider a Hamiltonian with a parametric dependence
$H(\varphi)$ such that it possesses a pair of zero eigenstates at some
$\varphi=\varphi_0$. Then the determinant of the matrix $H$
has a double zero at $\varphi_0$, i.e.  $ ({\pf H})^2 = \det H (\varphi) \propto
(\varphi-\varphi_0)^2$, thus $\pf H (\varphi) \propto
\varphi-\varphi_0$.  Any small perturbation of the Hamiltonian
(assuming it obeys the symmetry condition $\{\hat{H},\hat{\Xi}\}=0 $) can only shift the
value of the crossing point  $\varphi_0$, but its very existence
 is robust.  Apparently,  the above arguments are in  contradiction 
with well-known properties of Andreev levels in a usual
superconductive  quantum point, with energies given~\cite{Beenakker}
 by $E_{\pm} = \pm \Delta \sqrt{1-\mathcal{T}\sin^2\frac{\varphi}{2}}$, so that
 an arbitrary small reflection probability $r = 1- \mathcal{T}$ leads to the splitting
of zero levels present at $\varphi=\pi$ in the ballistic case
$\mathcal{T}=1$. 
The origin of this contradiction lies in the spin degeneracy present
in  usual systems  without spin-orbit interaction:
in the presence of such a degeneracy  the Pfaffian of the ballistic contact
 has a double zero, $\pf H(\varphi) \sim (\varphi-\pi)^2$, which is
 not robust to weak perturbations. Below we consider the generic case of
 strong spin-orbit coupling and thus no spin degeneracy.

 The sign of $\pf H$ changes simulteneously with  the
fermionic parity $F_{0}$ of the global ground-state of the system~\cite{Kitaev2000,Hopf}.
Indeed, the two eigenstates  $|e\rangle,|o\rangle$ of the total Hamiltonian,
 which become degenerate when $E_{e,o}=0$  at $\varphi=\varphi_0$, have fermionic numbers
differing by 1.  As the phase $\varphi$ passes $\varphi_0$, the roles of the ground state
 and the lowest excited state are interchanged.
 If the actual fermionic parity $F$ of the system cannot
 change due to conservation laws (which we will assume to be the
 case), we come to the following conclusion:
each time a pair of Majorana levels crosses $E=0$, the ground-state $|g_0\rangle$ is
transformed to the lowest \textit{excited} state $|e_0\rangle$, and vice
versa, $|e_0\rangle \to |g_0\rangle$.
We argue now that while the phase $\varphi$ changes on the $(0,2\pi)$
interval, an \textit{odd} number of such crossings occurs, i.e. after a $2\pi$
phase rotation our  system \textit{does not return} to its original state.

Consider first the  system shown in Fig.~\ref{fig:model} without the  cylindric channel but with 
two point vortices present in both superconductive films, on the upper
and lower surfaces. Each of them hosts a single MBS~\cite{FuKane2008}.
Due to finite thickness $L$ of the slab, these two MBS $\chi_{1,2}$
 are hybridized into a single complex fermion $\psi = \chi_1 +
 i\chi_2$. The energy $e_0(\varphi)$ of this fermionic mode is proportional, in
 general, to the amplitude of MBS tunnelling $\tilde{t} \sim e^{-L/\xi_{TI}}$ between the surfaces.
However, for $\varphi=\pi$ the tunnelling amplitude vanishes due to
destructive interference (see Supplement 1), thus a single level crossing
at $E=0$ occurs as $\varphi$ varies on the $(0,2\pi)$ period.
Let us now open the cylinder channel across the slab. It results in
a drastic increase of hybridization between upper and lower
superconductive surfaces, and in the appearence of an \textit{even} (due to Kramers degeneracy)
 number of  conductive modes. 
 Since these additional modes appear
in pairs only, the transformation $g_0 \to e_0$  occurs on 
each  $2\pi$-period of $\varphi$ variation. The above arguments prove
the existence of the anomalous
component $I_a(\varphi)$ of the Josephson current
which is odd under $2\pi$-shift. 
Below we calculate its magnitude and temperature dependence.

To find the current flowing along the hole
channel we first calculate the sub-gap spectrum of the contact.  We  
assume that the tube's radius $R$ and length $L$ (the latter coincides 
with the thickness of our TI slab) satisfy the conditions
\bqs
p_f^{-1}\ll R,L\ll \xi_0, \xi_{sc} 
\label{cond}
\eqs 
where $\xi_0= v_f/\Delta$,   $p_f$ is the Fermi momentum of surface
electrons  of the TI, and $\xi_{sc} = \sqrt{D/2\Delta} \ll \xi_0$ is the actual "dirty-limit"
coherence length in the superconductive film with diffusion coefficient $D$.
Inequlities \eqref{cond} mean, in particular, that we consider a short SNS-contact with
many transverse channels, $N_{ch} \sim p_fR$. To find the sub-gap spectrum we solve Bogolyubov-de
Gennes equations on both surfaces of the TI in the presence of the induced gap and
vortices, and match  obtained solutions with eigenmodes living on the inner cylindric
surface of the channel.

We use cylindric coordinates $r,\theta,z$ with the $z$-axis coinciding
with the tubes axis. The Hamiltonian \eqref{h1} in the presence of an Abrikosov vortex is:
\bqs
\hat{H}= v_f\mathbf{\sigma}\cdot (\mathbf{p}\tau_z-\nabla\theta/2)-
E_F\tau_z+\Delta(\mathbf{r})\tau_x\label{h bdg 2}
\eqs
The magnetic screening length is very long for thin films, and the flux  of the vortex is
distributed over a radius much greater than both $R$ and $\xi_{0}$, thus  we may neglect the 
vector-potential term in \eqref{h bdg 2}.
 For a fixed angular momentum $\nu$ we get
 $\Psi=e^{i\nu\theta-i\sigma_z\theta/2}\bmx u(r)\\ v(r)\emx $
 with radial wave functions $u(r)$ and $v(r)$ given by
\bqs
\bmx\hat{H}_{\nu-N/2}-\epsilon &\Delta(r)\\
\Delta(r)&-\hat{H}_{\nu+N/2}-\epsilon \emx
\bmx u\\v\emx=0 \label{eq bdg final}
\eqs
with $\hat{H}_m$ standing for
 $v_f \left(\sigma_x\left(p_r-\frac{i}{2r}\right)+\sigma_y\frac{m}{r}\right)-E_f$.
The profile of the gap function near the vortex center is $\Delta(r) \approx r\Delta/\sqrt{r^2+ 2\xi^2_{sc}}$.
 Equation \eqref{eq bdg final} can be solved for $\epsilon \ll\Delta$~\cite{Kopnin91}.
In the first order in $\epsilon/\Delta$ it yields:
\bqs
u= 
\left[c_1e^{-i\phi}w^{(1)}_{\nu-1/2}(r)
-ic_2e^{i\phi}w^{(2)}_{\nu-1/2}(r)\right]e^{-K}\label{u}\\
v = 
\left[-ic_1e^{i\phi}w^{(1)}_{\nu+1/2}(r)+
c_2e^{-i\phi}w^{(2)}_{\nu+1/2}(r)\right]e^{-K}\label{v}\\
K(r)=\int_0^r\Delta(\rho)d\rho\\ 
\phi(r)=e^{2K(r)}\int_r^\infty\left(\epsilon+\frac{\nu\Delta(\rho)}{p_f\rho}\right) e^{-2K(\rho)}d\rho \label{phi}\\
w^{(1,2)}_m(r)=e^{\pm\frac{im}{2}}\bmx H^{(1,2)}_{m-1/2}(p_fr)\\iH^{(1,2)}_{m+1/2}(p_fr)\emx\label{hankinsol}
\eqs
$H^{(1,2)}$ in~\eqref{hankinsol} denote Hankel functions.
 When $\nu,\epsilon=0$, solutions \eqref{u},\eqref{v} become 
exact~\cite{KhayKop}. 
Due to the $e^{-K}$ factor in Eqs.\eqref{u},\eqref{v}, the wavefunction
 $\Psi$ is localized in the vicinity of the tube within a length
$\sim \xi_0$.   Next we specify electronic eigenmodes in the channel. 
The Hamiltonian for a cylindrical surface has the form \cite{Zhang} 
\bqs
H_{c}=\left(\sigma_zp_z+\sigma_\theta p_\theta+\frac{i}{2R}\sigma_r\right)\label{h curc}
\eqs
with $\sigma_\theta=-\sigma_x\sin\theta+\sigma_y\cos\theta$ and $\sigma_r=\sigma_x\cos\theta+\sigma_y\sin\theta$.
 The eigenfunctions of \eqref{h curc} are
\begin{eqnarray}
\psi=e^{ipz}e^{i\theta\mu}e^{-i\sigma_z\theta/2}
\bmx\cos\frac{\alpha}{2}\\ i\sin\frac{\alpha}{2}\emx\label{sol cyl}
\end{eqnarray}
where $\alpha=\arctan\frac{\mu}{Rp}$. The energy spectrum for positive energies (counted from $E_F$)
 is $\epsilon_{\mu,\mathbf{p}}=\sqrt{p^2+\frac{\mu^2}{R^2}}$. 
The angular momentum is half-integer, $\mu=\frac12+n$, due to the Berry phase originating from the rotation of the spin.
We neglect the effect of the small magnetic flux $\Phi_R \ll \Phi_0$ inside the cylinder.
Note that the possibility to find conductive channels with no magnetic flux inside is due to the semiclassical condition
$p_f R \gg 1$; on the contrary, 
in the ultraquantum limit $p_f R \leq 1$ the presence of a $\Phi_0$ flux would be necessary to make the
 hole conductive, see Supplement to Ref.~\cite{Ostrovsky}.

Now we have to match wavefunctions \eqref{u},\eqref{v} on the upper and lower surfaces with wavefunctions \eqref{sol cyl} (and their  analog for the hole component with $\epsilon<0$) on the boundaries between the cylinder and flat surfaces. 
The proper matching condition for a sharp edge reads:
\bqs
\Psi_1 = \exp\left[-i\frac{\theta}{2}(\sigma\cdot\mathbf{n_\theta})\right] \Psi_2
\label{matching}
\eqs
Here $\mathbf{n}_\theta$ is the unit vector in the direction of the edge and $\theta$ is the rotation angle (equal to $\frac{\pi}{2}$
in our case). $\Psi_{1,2}$ are 4-component vector wavefunctions on both sides of the edge. The operator
in Eq.(\ref{matching}) rotates the spin by the angle $\theta$ around the direction $\mathbf{n}_\theta$.
 We provide an explicit form of the  matching equations and the derivation of the resulting energy spectrum
in the Supplements 2,3.

The resulting low-lying  levels with $\epsilon \ll \Delta$ as function of the phase difference
$\varphi$ are given by (up to the neglected terms of the order of $\delta\varepsilon \sim \nu\Delta/(p_fR)^2$):
\bqs
\epsilon_{\nu k} (\varphi) = \pm \varepsilon_0(\varphi-\pi-2\pi k)-\nonumber\\
-\varepsilon_1\left[\arcsin\frac{\nu}{p_fR}+\frac{\nu L}{2p_fR^2}\left(1-\frac{\nu^2}{(p_fR)^2}\right)^{-1/2}    \right]
\label{spectrum_low}
\eqs
where $\varepsilon_1 \approx \Delta$ and $\varepsilon_0 \approx \Delta/2$ for
 the case $\xi_{sc} \ll \xi_0$ we consider,  $|\nu| < p_f R$ and $k$ is an integer.
No propagating modes in the tube exist at $\nu>p_fR$, instead there is an exponentially
 small overlap between the bound states residing inside vortex cores on opposite surfaces. The energies of the
 global eigenstates
 with $|\nu| > p_fR$ are given by $\epsilon_\nu =\nu\frac{\Delta}{p_f\xi_0}\log\frac{\xi_0}{\xi_{sc}}$ up to exponentially weak
 $\varphi$-dependent corrections. 
 The result (\ref{spectrum_low}) is not applicable in the region $|\nu- p_fR|\sim (p_fR)^{1/3}$
 where the  crossover between hybridized and non-hybridized levels occurs.
\begin{figure}[tp]
	\centering
		\includegraphics[width=0.4\textwidth]{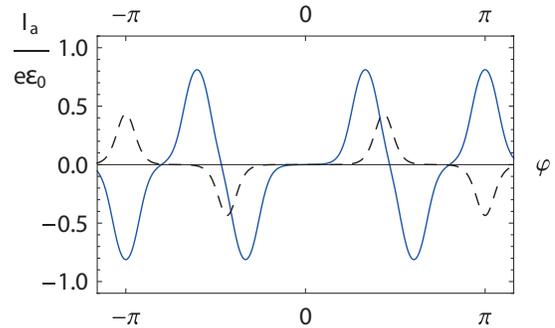}
	\caption{The anomalous current $I_a(\varphi)$  is computed for $p_fR = 2$ (dashed line) and $p_fR=3$ (blue line).
Other parameters are fixed as $p_f\xi=10, \quad p_f\xi_{sc}=5,\quad p_fL=6,\quad T=0.05\Delta$. 
The extrema of $I_a(\varphi)$ occur whenever a pair of conjugated levels crosses at $\epsilon=0$.}
\label{fig:long}
\end{figure}

\begin{figure}[tp]
	\centering
		\includegraphics[width=0.4\textwidth]{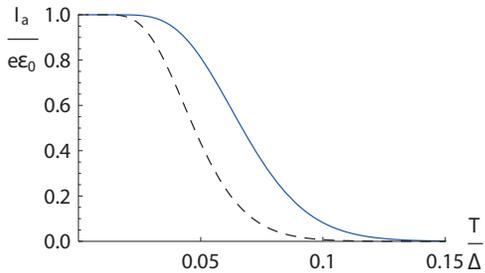}
	\caption{Temperature dependence of $I_a(\varphi=\pi)$ for the same sets of parameters as in Fig.\ref{fig:long}.
Blue and dashed lines correspond to $p_fR = 3$   and $p_fR=2$.}
	\label{fig:decay}
\end{figure}

The supercurrent through a short SNS-contact in thermal equilibrium 
can be expressed~\cite{Beenakker} in terms of Andreev levels: 
$I (\varphi) =-2e\sum_{j:\epsilon_j>0}\tanh\left(\frac{\epsilon_j}{2T}\right)\frac{\partial \epsilon_j}
{\partial\varphi}$.
Here all eigenstates are taken into account, regardless of their parity. The total current
 scales with the number of conductive 
channels in the hole: $I \sim e\Delta p_f R/\hbar$. 
To reveal any parity-related effects, we have to consider thermodynamic ensembles with odd and even
 numbers of quasiparticles separately. A standard route is to
 introduce  thermodynamic potentials $\Omega_{odd/even}$,
 describing odd and even numbers of quasiparticles correspondingly \cite{TuominenTinkham}.
Dividing the total current $I(\varphi)$ into a sum $I_n(\varphi)+I_a(\varphi)$ where $I_n(\varphi) $ 
is parity-independent, we obtain
 $I_a=(-1)^{F_0} e\frac{\partial}{\partial\varphi}(\Omega_{odd}-\Omega_{even})$, see Suppl.4. 
In terms of  the  spectrum of Andreev levels, it reads 
\bqs
I_a(\varphi) =(-1)^{F_0}\frac{2e f }{1-f^2}\sum_j
\frac{1}{\sinh\frac{\epsilon_j}{T}}\frac{\partial \epsilon_j}{\partial\varphi}
\label{curn2}
\eqs 
where $f =\prod_j\tanh\frac{\epsilon_j}{2T} = f_{hyb}\cdot f_{non}$ and both the products and the sum
 are done over levels with $\epsilon_j>0$. 
Factors $f_{hyb}$ and $f_{non}$  correspond to the hybridized (current-carrying) and non-hybridized levels.
The physical meaning of $I_a$ is the current difference between an odd and an even state of the system.
Eqs. (\ref{spectrum_low},\ref{curn2}) constitute our major quantitative result. Two examples of the
 $I_a(\varphi)$ dependence
computed using Eqs.(\ref{spectrum_low},\ref{curn2}) are presented in Fig.~\ref{fig:long}.
$I_a(\varphi)$ experiences $\approx 2p_fR$  oscillations in the $(-\pi,+\pi)$ interval  and
 has opposite signs at $\varphi = \pm \pi$ where the amplitude of $I_a(\varphi)$ is  maximal. 
Note that approximations used to derive Eq.(\ref{spectrum_low}) may lead to a deformation of the
$I_a(\varphi)$ dependence, leading to an inhomogeneous shift of its oscillating pattern
by the amount $\sim (p_fR)^{-1}$. It should not affect, however, the maximal value $I_a^{\rm max}$ of 
the anomalous current. The temperature dependence of $I_a^{\rm max}$ is presented in Fig.~\ref{fig:decay}.

To analyze the temperature dependence of $I_a(\varphi)$ we consider Eq.(\ref{curn2}) in several limits.
In the range of  $T$  much higher than the typical level spacing $\varepsilon_1/p_fR$, the subproduct $f_{hyb} \sim e^{-\pi^2T p_fR/4\varepsilon_1}$ (for the derivation of this and the following formulae  see Supplement 4). If, in addition, $T \gg E_0\equiv\Delta\frac{R}{\xi_0}\ln\frac{\xi_0}{\xi_{sc}}$, then the subproduct $f_{non}$
is also small: $f_{non} = e^{-\pi^2T/2\delta}$, where $\delta = (\Delta/p_f\xi_0)\ln\frac{\xi_0}{\xi_{sc}}$ is the typical  level spacing within a single vortex core. Thus  $I_a$ decays
exponentially with temperature at $ T > T_1 = 
\min(\frac{\Delta}{p_fR} ,\frac{\Delta R}{\xi_0}\ln\frac{\xi_0}{\xi_{sc}} ) $.
At lower temperatures $T\ll T_1$
we find $f_{hyb} = \tanh(\epsilon_1/2T)$, where $\epsilon_1(\varphi)$ is
the lowest hybridized level, therefore
\begin{eqnarray}
\label{curn3}
I_a(\varphi)
 \simeq\frac{ e\varepsilon_0\cdot f_{non}}{\cosh^2\frac{\epsilon_1}{2T}- f_{non}^2 \sinh^2\frac{\epsilon_1}{2T}}
\\ \label{f_non}
f_{non} = \exp\left[-\frac{4e^{-E_0/T}}{1-e^{-\delta/T}}\right]
\end{eqnarray}
Simple analysis of Eqs.(\ref{curn3},\ref{f_non}) leads to  the second characteristic temperature 
$T_2 = E_0/\ln(p_f\xi_0)$. Depending on parameters,
$T_2$ may be both higher and lower than $T_1$. In addition, we mention the existence of
non-hybridized subgap states localized in superconductive films near vortex cores, which may lead to some
 suppression
of the crossover temperature $T_2$; however, we do not expect their effect to be drastic.
Finally, 
the usual parity-effect temperature $T_3 = \Delta/\ln(\nu V\Delta)$, see Ref. ~\cite{TuominenTinkham},
 puts an additional restriction
 for the temperature region where an anomalous current could be observed.

Summarizing the above analysis, we find that for the anomalous current $I_a$ to be detectable,
 the following condition must be met:
\bqs
T \leq \min\left(\frac{\Delta}{p_fR},\, \frac{2\Delta R\ln(\xi_0/\xi_{sc})}{\xi_0 \ln(p_f\xi_0)}, \,\frac{\Delta}{\ln(\nu V\Delta)}\right)
\label{t2}
\eqs
The temperature dependence of the anomalous current is shown in Fig.\ref{fig:decay} for a specific 
choice of parameters such that
 $T_2<T_1$.


To conclude, we proposed a setup using strong topological insulator covered by superconductive films,
which allows the detection of Majorana bound states through the measurement of an 
anomalous $4\pi$-periodic component $I_a$ of the Josephson current. The temperature dependence of 
the $I_a$ amplitude is calculated, and the conditions for the proposed effect to be observed are found.

We are grateful to L. B. Ioffe, D. A. Ivanov, A. Yu. Kitaev, J. E. Moore  and  P. M. Ostrovsky 
for numerous discussions and advises. 
This research was supported by the RFBR grant \# 10-02-00554
and by the RAS program "Quantum physics of condensed matter".


\section{Online supplmentary material}

\subsection{1. Degeneracy of MBS at $\varphi=\pi$}
Beside of the $\Xi$-symmetry inherent to Bogolyubov-de Gennes equation, our model has a symmetry that is the composition $U$ of rotation $R$ and time-reversal $\Theta$. Indeed, by rotating our system around an axis lying in the $z=0$-plane (the plane lying between the surfaces of our slab), we arrive at the same system as by conjugating $\Delta$, possibly with a phase shift. Selecting a gauge, in which the superconducting phase is given by $\arg\Delta(\theta,z=L/2)=\theta=\arg\Delta(\theta,z=L/2)-\varphi$, we find that $U$ commutes with the BdG Hamiltonian if the axis of rotation forms a $\varphi/2$ angle with the $x$-axis. The part of $R$ that acts on spin variables  equals $-i[\sigma_x\cos\varphi/2+\sigma_y\sin\varphi/2]$, while $\Theta=i\sigma_yK$, where $K$ stands for complex conjugation. Hence, $U\sim \sin\varphi/2+i\sigma_z\cos\varphi/2$. The first term commutes with $\Xi\equiv\sigma_y\tau_yK$, while the second term does not, in general. We see that $U$ commutes with $\Xi$ only if $\varphi=\pi$. The commutation relations $[H,U]=0;[U,\Xi]=0;\{H,\Xi\}=0$ lead to the degeneracy of the $E=0$ levels of
our system at $\varphi=\pi$.

\subsection{2. The matching equations}
Consider two surfaces -- 1 and 2 -- joining on a line at the angle of $\theta_0$. Let $\Psi_{1,2}$ be the wave function on the edge of surface 1 and 2 respectively. We replace the sharp boundary by a smooth cylindric transition with a radius  $r$. We use the natural cylindric coordinates of this cylindric sector with $\theta=0$ at its boundary with surface 1. Next we decompose $\Psi_{1,2}$ into the set of functions  (11): 
\bqs
\Psi_1=\sum_{p,\mu}a_{p,\mu}\Psi_{p,\mu}\\
\Psi_2=\sum_{p,\mu}a_{p,\mu}e^{i\theta\mu-i\sigma_z\theta_0/2}\Psi_{p,\mu}
\eqs
with $\Psi_{p,\mu}=e^{ipz}\bmx\cos\frac{\alpha}{2}\\i\sin\frac{\alpha}{2}\emx$.
In the limit $r\to0$ we have to choose $\mu =0$ since $\sqrt{p^2+\frac{\mu^2}{r^2}}=\epsilon_{p,\mu}=const$. This allows us to write $\Psi_2=e^{-i\sigma_z\theta_0/2}\Psi_1$. The basis-independent form of this equation is given by Eq.(12).

\subsection{3. The spectrum equations}
The set of equations defining our systems spectrum consists of 8 equations - two sets of 4 matching equations corresponding to the two ends of the cylindric hole. To derive these equations, let we first write down the  wave functions for some fixed momentum $\nu$; we will do it in linear approximation in  small parameter $\frac{1}{p_fR} \ll 1$.  The fixed angular momentum $\nu$ on the surfaces corresponds to  $\nu_\pm=\nu\pm1/2$ in the channel: electron waves have $\mu=\nu_-$, while hole waves have $\mu=\nu_+$. After we have chosen the proper angular momenta, we only need to match wave functions at some fixed angle, say $\theta=0$. The w.f. in the cylinder is a superposition of four waves, representing electrons and holes propagating up and down the tube: $\Psi_{cyl}=\bmx a_\uparrow u_\uparrow+a_\downarrow u_\downarrow & b_\uparrow v_\uparrow+b_\uparrow v_\uparrow\emx^T$.

\bqs
u_\uparrow=e^{i(p_f+q)z\cos\alpha_-}\begin{pmatrix}\cos\alpha_-/2\\ i\sin\alpha_-/2\end{pmatrix}\\
u_\downarrow=e^{-i(p_f+q)z\cos\alpha_-}\begin{pmatrix}-i\sin\alpha_-/2\\ \cos\alpha_-/2\end{pmatrix}\label{sup cyl1}\\
v_\downarrow=e^{i(p_f-q)z\cos\alpha_+}\begin{pmatrix}\cos\alpha_+/2\\ i\sin\alpha_+/2\end{pmatrix}\\
v_\uparrow=e^{-i(p_f-q)z\cos\alpha_+}\begin{pmatrix}-i\sin\alpha_+/2\\ \cos\alpha_+/2\end{pmatrix}\label{sup cyl2}\\
\alpha_\pm=\arcsin\frac{\nu\pm1/2}{p_fR}\qquad\qquad q\equiv \frac{\epsilon}{v_f}
\eqs  
Next we write out the surface wave functions. At $r\ll \xi$ the solution \eqref{hankinsol} transforms into 
\bqs
u=c_1e^{-i\phi_0}w^{(1)}_{\nu-1/2}(r_+)-ic_2e^{i\phi_0}w^{(2)}_{\nu-1/2}(r_+)\\
v=-ic_1e^{i\phi_0}w^{(1)}_{\nu+1/2}(r_-)+c_2e^{-i\phi_0}w^{(2)}_{\nu+1/2}(r_-)
\eqs
with $r_\pm=r(1\pm q/p_f)$ and $\phi_0=\phi(0)\simeq q\xi_0/2+\frac{\nu}{p_f\xi_0}\log\frac{\xi_0}{\xi_{sc}}$ in the dirty limit $\xi_{sc}\ll\xi_0$.
Using the asymptotics of Hankel functions with large arguments, we find the w.f. on the upper surface at $r=R, z=L, \theta=0$,
within $\sim 1/p_fR$ accuracy (we neglect terms $\sim (p_fR)^{-2}$):
\bqs
\Psi_{up}(R)=C_1\bmx  \exp[i\beta-2i\phi_0+\varphi]\\i\exp[-2i\phi_0+\varphi]\\1\\i\exp[-i\beta] \emx+\nonumber\\+C_2\bmx  \exp[-i\beta+2i\phi_0+\varphi]\\i\exp[2i\phi_0+\varphi]\\1\\i\exp[i\beta] \emx
\eqs
with some coefficients $C_1,C_2$ and $\cos\beta=\nu/p_fR$. This formula is correct unless $|\nu-p_fR|\lesssim (p_fR)^{1/3}$. $\Psi_{down}(R)$ is obtained by introducing two new coefficients $Q_1,Q_2$ instead of $C_1,C_2$ and putting $\varphi=0$.

Now we can construct the matching equations according to the matching rule (12). Acting on $\Psi_{cyl}(z=L)$ and $\Psi_{cyl}(z=0)$ by $\frac{1}{\sqrt2}(1-i\s_y)$ and $\frac{1}{\sqrt2}(1+i\s_y)$ respectively we finally get

\begin{widetext}
\begin{small}
\bqs
\label{fullset}
A_\uparrow e^{i(p_f+q)L\cos\alpha_-}\bmx 1\\e^{i\alpha_-}\emx +A_\downarrow e^{-i(p_f+q)L\cos\alpha_-}\bmx -e^{i\alpha_-}\\1\emx &=& C_1\bmx\exp[i\beta-2i\phi_0+\varphi]\\i\exp[-2i\phi_0+\varphi] \emx+C_2\bmx  \exp[-i\beta+2i\phi_0+\varphi]\\i\exp[2i\phi_0+\varphi]\emx\\ \nonumber
B_\downarrow e^{i(p_f-q)L\cos\alpha_+}\bmx 1\\e^{i\alpha_+}\emx +B_\uparrow e^{-i(p_f-q)L\cos\alpha_+}\bmx -e^{i\alpha_+}\\1\emx
&=&
C_1\bmx1\\i\exp[-i\beta] \emx+C_2\bmx 1\\i\exp[i\beta]\emx\\ \nonumber
A_\uparrow \bmx e^{i\alpha_-}\\-1\emx +A_\downarrow \bmx 1\\e^{i\alpha_-}\emx
&=&
Q_1\bmx\exp[i\beta-2i\phi_0]\\i\exp[-2i\phi_0] \emx+Q_2\bmx  \exp[-i\beta+2i\phi_0]\\i\exp[2i\phi_0]\emx\\ \nonumber
B_\downarrow \bmx e^{i\alpha_+}\\-1\emx +B_\uparrow \bmx 1\\e^{i\alpha_+}\emx
&=&
Q_1\bmx1\\i\exp[-i\beta] \emx+Q_2\bmx 1\\i\exp[i\beta]\emx\\ \nonumber
\eqs
\end{small}
\end{widetext}
Since $\alpha_\pm = \pi/2-\beta$ within our approximation, the spinors with coefficients  $A_\uparrow$ and $C_1$ in the first equation of the system (\ref{fullset}) are collinear. The same is true for the spinors with coefficients  $A_\downarrow$ and $C_2$ in the same equation. A similar statement is true for each of the other Eqs.(\ref{fullset}). Consequently, the system 
(\ref{fullset}) splits into two simple subsystems:
\begin{small}
\bqs
\label{set1}
A_\uparrow &= & C_1\exp[i\beta-2i\phi_0+i\varphi-i(p_f+q)L\cos\alpha_-]\\ \nonumber
B_\downarrow &= &C_1\exp[-i(p_f-q)L\cos\alpha_+]\\ \nonumber
A_\uparrow & = & Q_2\exp[-i\pi/2+2i\phi_0]\\ \nonumber
B_\downarrow &= & Q_2\exp[-i\pi/2+i\beta] \nonumber
\eqs
\end{small}
and
\begin{small}
\bqs
\label{set2}
A_\downarrow &=&C_2\exp[i\pi/2+2i\phi_0+i\varphi+i(p_f+q)L\cos\alpha_-]\\ \nonumber
B_\uparrow &=& C_2\exp[i\pi/2+i\beta+i(p_f-q)L\cos\alpha_+]\\ \nonumber
A_\downarrow & = &Q_1\exp[-2i\phi_0+i\beta]\\ \nonumber
B_\uparrow &= &Q_1 \nonumber
\eqs
\end{small}
These equations give the following set of the two spectral equations
\bqs
4\phi_0=-2\alpha\pm(\varphi-\pi-2\pi k)+\nonumber\\+(p_f-q)L\cos\alpha_+-(p_f+q)L\cos\alpha_-	\label{spectrum1}
\eqs
where the upper/lower signs refer to the systems (\ref{set1}) and (\ref{set2}) correspondingly. Since we assume $L\ll\xi_0$ and $R\ll\xi$, we neglect the $\sim qL$ term in the r.h.s of \eqref{spectrum1} and the $\nu$-dependent term in $\phi_0$, arriving at the final formula (13).

\subsection{4. The anomalous current}

The thermodynamic potential of a system with a fixed parity can be written as
\bqs
\Omega_{odd/even}=-T\ln\frac{\prod_i\left(1+e^{-\beta \epsilon_i}\right)\mp \prod_i\left(1-e^{-\beta \epsilon_i}\right)}{2}
\eqs
with $\beta=T^{-1}$. The difference $\delta\Omega=\Omega_{odd}-\Omega_{even}$ equals
\bqs
\delta\Omega=-T\ln\frac{1-f}{1+f}\qquad\text{with}\qquad
f\equiv\prod_i\tanh\frac{\beta \epsilon_i}{2}
\eqs
Hence
\begin{widetext}
\bqs
I_a=e\delta\Omega'_\varphi=\frac{2Tef'_\varphi}{1-f^2}=\frac{2Tfe}{1-f^2}\sum_i\frac{\beta \epsilon'_{i,\varphi}}{2\cosh^2\frac{\beta \epsilon_i}{2}\tanh\frac{\beta \epsilon_i}{2}}=\frac{2fe}{1-f^2}\sum_i\frac{\epsilon'_{i,\varphi}}{\sinh\beta \epsilon_i}.
\eqs
\end{widetext}

It is useful to divide $f$ into factors corresponding to the hybridized and non-hybridized parts of the discrete spectrum.
\bqs
f=\prod_{|\nu|<p_fR}\tanh\frac{\epsilon_\nu}{2T}\prod_{|\nu|>p_fR}\tanh\frac{\epsilon_\nu}{2T}=f_{hyb}f_{non}
\eqs
We neglect the exponentially small dependence of non-hybridized energies on $\varphi$, so that
\bqs
I_a=\frac{2ef_{hyb}f_{non}}{1-f^2_{hyb}f^2_{non}}J\qquad\qquad J=\sum_{|\nu<p_fR|}\frac{\partial \epsilon_{\nu}/\partial\varphi}{\sinh(\epsilon_\nu/T)}\label{sup ia}
\eqs
First let us analyze $f_{hyb}$. The typical level spacing is $\omega_0 \sim \frac{\Delta}{p_fR}$. If $T\gg\omega_0$, we can write $f_{hyb}=\exp\left[\int_0^\infty \ln\tanh(\frac{\omega_0\nu}{2T})d\nu \right]=\exp\left[-\frac{\pi^2 T}{4\omega_0} \right]$. For small $L/R$ and $\varphi=\pi$ this gives $\exp\left[-\frac{\pi^2 p_fR T}{2\Delta} \right]$    In the opposite case $T\ll\omega_0$ we can write $f_{hyb}=\tanh{\frac{E_h}{2T}}$, where $E_h$ is the lowest hybridized energy level.

Next we consider factor $J$. At $T\gg\omega_0$ we use $\sinh(\epsilon/T)\simeq
\epsilon/T$ at small energies and get
$J\lesssim\sum_{n=1}^{T/\omega_0}\frac{T\Delta}{n\omega_0}=\frac{T\Delta}{\omega_0}\log\frac{T}{\omega_0}$.
At $T\ll\omega_0$ the amplitude of $J$ is dominated by the lowest
hybridized level $\epsilon_1$ and
we get $|J|=\frac{\Delta}{2\sinh{\epsilon_1/T}}$. 

Finally, we analyze $f_{non}$. The spectrum of the non-hybridized de Gennes states is described by $E_{non}=\frac{\Delta}{p_f\xi}\ln\left(\frac{\xi_0}{\xi_{sc}}\right)\left[p_fR+n\right]=E_{0}+n\delta$ with $n=0,1,2...$. There are two copies of these series -- one for the upper and one for the lower surface.
If $(T-E_0)\gg \delta$ we can rewrite 
\begin{widetext}
\bqs
f_{non}=\exp\left[2\int_{p_fR}^\infty \ln\tanh\frac{n\delta}{2T}dn   \right]
=\exp\left[-\frac{\pi^2T}{2\delta}-\frac{4T}{\delta}\int_0^{\frac{E_0}{2T}}\ln\tanh xdx  \right]
\eqs
The second term can be neglected, if $T\gg E_0$. At low temperatures $T\ll E_0$ we have
\bqs
f_{non}=\exp\left[2\sum_n \ln[1-2e^{-\frac{E_n}{T}}]   \right]=
\exp\left[-4e^{-\frac{E_0}{T}}\sum_{n=0}^\infty e^{-\frac{n\delta}{T}}\right]=
\exp\left[-4\frac{e^{-\frac{E_0}{T}}}{1-e^{-\frac{\delta}{T}}}\right]\label{sup non low}
\eqs
The low-temperature expressions derived for $f_{hyb}, f_{non}, J$ lead to formulae (15,16). 
\end{widetext}


\begin{thebibliography}{99}

\bibitem{FuKane2008}  
L. Fu and C. L. Kane, Phys.Rev.Lett. \textbf{100}, 096407 (2008).

\bibitem{MajoInDefects}  
M. Wimmer, A.R. Akhmerov et al Phys. Rev. Lett. \textbf{105}, 046803 (2010)

\bibitem{SemiNanowire}  
J. D. Sau, S. Das Sarma et al arXiv:1006.2829 (2010)

\bibitem{BeenakkerAkhmerovNilsson2009}  
A. R. Akhmerov, J. Nilsson, and C. W. J. Beenakker arXiv:0903.2196 (2009)

\bibitem{Lee2009}  
K. T. Law, P. A. Lee, and T.K. Ng Phys. Rev. Lett. \textbf{103}, 237001 (2009)

\bibitem{Sarma1}  
R. M. Lutchyn, J. D. Sau, and S. Das Sarma,  Phys. Rev. Lett. \textbf{105}, 077001 (2010)

\bibitem{Sarma2}  
J. D. Sau, S. Tewari, and S. Das Sarma,  arXiv:1004.4702 (2010)

\bibitem{Kitaev2000}   A. Kitaev, arXiv: 0010440v2, (2000)

\bibitem{Hopf}   Y. Ran, P. Hosur, and A. Vishwanath,  arXiv:1003.1964 (2010)

\bibitem{Kopnin91}   N. B. Kopnin and M. M. Salomaa, Phys. Rev. B \textbf{44}, 9667 (1991).

\bibitem{KhayKop}   I. M. Khaymovich, N. B. Kopnin et al Phys. Rev. B \textbf{79}, 224506 (2009)

\bibitem{Zhang}   Y. Zhang, Y. Ran, and A. Vishwanath Phys. Rev. B \textbf{79}, 245331 (2009)

\bibitem{Ostrovsky}  P. M. Ostrovsky, I. V. Gornyi, and A. D. Mirlin, Phys. Rev. Lett.
\textbf{105}, 036803 (2010).

\bibitem{Beenakker}   C. W. J. Beenakker and H. van Houten, Phys. Rev. Lett. \textbf{66}, 23 (1991) 

\bibitem{TuominenTinkham}   M. T. Tuominen, M. Tinkham et al, Phys. Rev. Lett. \textbf{69}, 1997 (1992).



\end{thebibliography}
\end{document}